%% file: main_v3.tex
\documentclass[aps,prl,preprint,superscriptaddress]{revtex4-2}

\usepackage{color}
\usepackage{graphicx}
\usepackage{amsmath}

\begin{document}
\preprint{OU-HET-1295}

\title{Probing Internal Conversion and Dark-Matter--Induced De-excitation of $^{180\mathrm{m}}$Ta with a $\gamma$-ray TES Array}

\newcommand{\tohoku}{\affiliation{Research Center for Neutrino
    Science, Tohoku University, Sendai, Miyagi 980-8578, Japan}}
\newcommand{\rcnp}{\affiliation{Research Center for Nuclear Physics, 
    The University of Osaka, Ibaraki, Osaka 567-0047, Japan}}
\newcommand{\obihiro}{\affiliation{Department of Human Science, Obihiro University of Agriculture and Veterinary Medicine, Obihiro, Hokkaido 080-8555, Japan}}
\newcommand{\aist}{\affiliation{National Institute of Advanced Industrial Science and Technology (AIST), Tsukuba, Ibaraki 305-8565, Japan}}
\newcommand{\osaka}{\affiliation{Department of Physics, The University of Osaka, Toyonaka, Osaka 560-0043, Japan}}
\newcommand{\rikkyo}{\affiliation{Rikkyo}}

\author{A.~Gando}\tohoku\obihiro
\author{K.~Ichimura}\tohoku
\author{K.~Ishidoshiro}\tohoku
\author{T.~Kikuchi}\aist
\author{T.~Kishimoto}\rcnp
\author{A.~Takeuchi}\tohoku
\author{R.~Sato}\osaka
\author{R.~Smith}\aist




\begin{abstract}
We propose and evaluate a source-equals-detector search for the de-excitation of the long-lived isomer $^{180\mathrm{m}}\mathrm{Ta}$ in natural tantalum (Ta), using a $\gamma$-ray transition-edge-sensor (TES) array. 
Two capabilities absent in conventional high-purity germanium (HPGe)
searches are exploited: (i) near-unity containment of low-energy
secondaries (internal-conversion~(IC) electrons and characteristic X rays)
and of the nuclear recoil, enabling a calorimetric, event-by-event
measurement of the total deposited energy in the absorber; and
(ii) a delayed-coincidence tag using the subsequent ${}^{180}$Ta
electron-capture (EC) decay to ${}^{180}$Hf.
We evaluate the $3\sigma$ discovery reach for IC
and for dark-matter~(DM)–induced de-excitation in two benchmark scenarios:
a strongly interacting DM subcomponent and inelastic DM with
off-diagonal couplings.
Using a background model based on intrinsic radioactivity in the Ta
absorber and realistic detector performance, we show that arrays with
$N_{\mathrm{TES}} = 256$ and $1,000$ pixels can reach the theoretically
expected IC half-life within $2.6$~yr and $0.66$~yr, respectively.
For an array with $N_{\mathrm{TES}} = 10^4$ and a five-year exposure,
the projected sensitivity to DM-induced de-excitation surpasses limits
inferred from HPGe non-observations of ${}^{180\mathrm{m}}$Ta and probes
regions of parameter space not covered by current direct-detection
experiments.
\end{abstract}


\maketitle

\section{Introduction}
The long-lived isomeric state of tantalum-180, $^{180\mathrm{m}}\mathrm{Ta}$, is the only known naturally occurring nucleus whose isomeric state ($E_x\simeq77.2~\mathrm{keV}$, $J^\pi=9^-$) outlives the ground state ($J^\pi=1^+$). 
This extreme metastability is protected by a combination of $K$-forbiddenness, spin forbiddenness, and a parity change, resulting in an exceptionally suppressed transition probability to the ground state~\cite{Auerbach2017,Ejiri:2017dro}. Despite several decades of experimental effort, no decay of $^{180\mathrm{m}}$Ta has been observed to date~\cite{Bauminger1958,Ryves1980,Cumming:1985zza,Hult:2006tp,Lehnert:2016iku,Majorana:2023ecz,Cerroni:2023qoo}. Existing underground measurements provide only lower limits on the half-life. 

Several decay modes of $^{180\mathrm{m}}$Ta are theoretically allowed: 
(i) electron capture (EC) to $^{180}$Hf ($J^\pi=6^+$), 
(ii) $\beta^-$ decay to $^{180}$W ($J^\pi=6^+$),
(iii) de-excitation~($\gamma$-ray emission and internal conversion (IC)) to the $J^\pi=2^+$ state and then to the $J^\pi=1^+$ state, and (iv) possibly $\alpha$ decay to $^{176}$Lu, though strongly suppressed. 
In addition, theoretical work has pointed out that interactions with dark matter (DM) can induce de-excitation of $^{180\mathrm{m}}$Ta that bypasses the usual selection rules~\cite{Pospelov:2019vuf,Smirnov:2024jvj}. 

A measurement of the $^{180\mathrm{m}}$Ta decay would have implications
in several areas of nuclear and astroparticle physics. 
First, it provides a stringent benchmark for nuclear-structure calculations on high-spin, multi-particle configurations and strongly hindered electromagnetic transitions~\cite{Auerbach2017,Ejiri2017a}.
Second, the isomer plays a special role in nucleosynthesis: the solar abundance of $^{180\mathrm{m}}$Ta is thought to arise from a combination of $s$-process branchings, $\gamma$-process (or $p$-process) photodisintegration in explosive O/Ne burning, and the $\nu$-process in core-collapse supernovae~\cite{Mohr:2006wn,Hayakawa2010,Roberti2023,Malatji2019}.
The survival fraction of $^{180\mathrm{m}}$Ta at freeze-out is controlled by interlevel couplings through intermediate states and by the vacuum lifetime of the isomer, both of which remain uncertain. An experimental determination (or a stringent limit) of the $^{180\mathrm{m}}$Ta half-life would therefore provide valuable input to reaction-network calculations and the thermal-equilibration picture~\cite{Mohr:2006wn,Belic:2002sq}.
Third, DM-induced de-excitation of nuclear isomers offers a novel probe of DM that is complementary to conventional direct-detection searches. 
In scenarios with a strongly interacting DM subcomponent, nuclei can act as ``accelerators'' that convert the stored excitation energy into detectable nuclear recoils~\cite{Pospelov:2019vuf}. Inelastic DM with off-diagonal couplings provides another mechanism for inducing nuclear transitions~\cite{Tucker-Smith:2001myb}. For $^{180\mathrm{m}}$Ta, such processes would populate the low-lying
$J^\pi = 2^+$ and $1^+$ states, leading to de-excitation signatures in the 40--80~keV range together with a characteristic recoil spectrum. 

Most experimental searches have relied on high-purity germanium (HPGe) detectors operated in underground laboratories~\cite{Cumming:1985zza,Hult:2006tp,Lehnert:2016iku,Majorana:2023ecz,Cerroni:2023qoo}. These searches have also been reinterpreted as constraints on DM interactions~\cite{Lehnert:2019tuw,Majorana:2023ecz}. However, HPGe detectors are intrinsically insensitive to low-energy secondaries such as IC electrons and characteristic X rays. As a result, they cannot distinguish standard IC de-excitation from DM-induced de-excitation on an event-by-event basis, and the ultimate reach of HPGe-based searches is limited by the IC half-life itself.

Lehnert et al.~\cite{Lehnert:2019tuw} have briefly pointed out that transition-edge sensor (TES)-based microcalorimeters~($\gamma$-TES) with a source-equals-detector configuration could provide a promising path for future searches for $^{180\mathrm{m}}$Ta. 
A similar idea was discussed using metallic magnetic calorimeters (MMCs)~\cite{Smirnov:2024jvj}. In this work, we develop this $\gamma$-TES concept into a quantitative sensitivity study for $^{180\mathrm{m}}$Ta de-excitation. The use of massive Ta as a $\gamma$-TES  absorber has already been demonstrated~\cite{Irimatsugawa2015,ohno2017}. A Ta absorber with dimensions of order $(1.5~\mathrm{mm})^3$ can host a sizeable number of $^{180\mathrm{m}}$Ta nuclei while maintaining sub-keV energy resolution. Assuming natural tantalum with an isomeric abundance of 0.012\%~($1.2\times 10^{-4}$), a (1.5~mm) Ta absorber contains $N_T \simeq 2.3\times10^{16}$ $^{180m}$Ta nuclei per pixel, i.e. $5.8\times10^{18}$ nuclei for 256 pixels and $2.3\times10^{20}$ nuclei for 10,000 pixels.

The calorimetric nature of $\gamma$-TES devices provides two key capabilities that are absent in HPGe searches. 
First, the detector measures the total energy deposited in the absorber, including IC electrons and the accompanying atomic-relaxation products (characteristic X rays and Auger electrons), thereby providing a prompt handle on the de-excitation of $^{180\mathrm{m}}$Ta even when it proceeds predominantly via IC. For Ta, the characteristic fluorescence associated with L-shell IC lies at $\sim 8$--$12$~keV (L lines). Using the NIST mass attenuation coefficients for Ta, the attenuation length is $\lambda \simeq 3.6~\mu$m at 8~keV, which is far smaller than the mm-scale absorber dimension; thus X-ray escape is strongly suppressed except for events occurring very near the surface. In addition, for DM-induced de-excitation the associated nuclear recoil can also be recorded in the same calorimetric measurement, adding a distinct low-energy component that allows IC- and DM-induced channels to be separated based on their energy distributions. Second, the subsequent electron-capture (EC) decay of $^{180}$Ta can be recorded directly in the same source-equals-detector absorber via the atomic relaxation energy in the daughter $^{180}$Hf. Crucially, this delayed signature is present whether the EC populates the $^{180}$Hf ground state or an excited state: HPGe searches are primarily sensitive only to the de-excitation $\gamma$ rays from the excited-state branch, whereas a calorimetric $\gamma$-TES measures the EC atomic-relaxation energy in either case (with any subsequent nuclear de-excitation energy contributing additionally when applicable). This enables a robust delayed-coincidence tag between the prompt $^{180\mathrm{m}}$Ta de-excitation and the delayed $^{180}$Ta EC decay.

In this paper we quantify the discovery reach of a $\gamma$-TES array
for IC and DM-induced de-excitation of $^{180\mathrm{m}}$Ta in a
source-equals-detector configuration.
We construct a signal model that includes IC, strongly interacting DM,
and inelastic DM, together with a delayed tag on the subsequent EC of
$^{180}$Ta. 

Intrinsic radioactivity in the Ta absorber is taken as the dominant
background, and we develop a background model that incorporates both
accidental and correlated components.
Using the Asimov approximation for discovery significance, we evaluate
the $3\sigma$ reach as a function of array size and observation time.
For benchmark arrays with $N_{\mathrm{TES}} = 256$ and 1\,000 pixels,
the IC half-life sensitivity approaches the theoretical expectation
within observation times of order a few years.
For a larger array with $N_{\mathrm{TES}} = 10^4$ and a five-year
exposure, the projected discovery reach for DM-induced de-excitation
surpasses bounds inferred from HPGe non-observations of
$^{180\mathrm{m}}$Ta and probes regions of parameter space not covered
by current direct-detection experiments.

\section{Low-temperature microcalorimeters}
Various types of low-temperature microcalorimeters have been developed, including Superconducting Tunnel Junctions~(STJs), Microwave Kinetic Inductance Detectors~(MKIDs), neutron transmutation doped~(NTD) thermistors, and MMCs. In this study, we focus on $\gamma$-TES, because of their maturity for multiplexed readout and demonstrated performance with massive absorbers. Moreover, a $\gamma$-TES using Ta as a $\gamma$-ray absorber has already been developed by the National Institute of Advanced Industrial Science and Technology (AIST) and the University of Tokyo~\cite{Irimatsugawa2015}. 

In prior work, a Ta absorber of $0.5\times0.5\times0.3$~mm$^3$ was mounted on a TES via Au stub attachment, demonstrating feasibility for $\gamma$-ray detection. The best achieved FWHM energy resolution was $\Delta E_\mathrm{FWHM}=465$~eV~\cite{ohno2017}. 
More recently, a $\gamma$-TES array with massive absorbers on a thick SiO$_2$/Si$_x$N$_y$/SiO$_2$ membrane has been fabricated~\cite{kikuchi2021}, enabling multi-pixel configurations suitable for source-equals-detector configurations, by some of the present authors. 

Here we assume a per-pixel energy resolution extrapolated from previous measurements using $\Delta E_\mathrm{res} \propto \sqrt{C} \propto \sqrt{M}$, where $C$ is the heat capacity and $M$ is the absorber mass. We also assume a Ta absorber of cubic geometry with edge length 1.5~mm—the largest size currently feasible on the thick SiO$_2$/Si$_x$N$_y$/SiO$_2$ membrane.

\section{Signal model}
We exploit the $\gamma$-TES capabilities together with a delayed-coincidence technique to search for de-excitation of $^{180\mathrm{m}}$Ta. As prompt channels we consider: (i) IC, (ii) de-excitation induced by strongly interacting DM, and (iii) de-excitation induced by inelastic DM. Pure $\gamma$-ray de-excitation is expected to be far slower than IC for the transition of interest~\cite{Ejiri:2017dro} and is neglected. As delayed events we consider the subsequent EC of $^{180}$Ta to $^{180}$Hf with a half-life of 8.15~hr. For the DM-induced de-excitation, we follow the formalism of Ref.~\cite{Pospelov:2019vuf} for a direct comparison with previous constraints derived
from HPGe searches~\cite{Majorana:2023ecz}. Related treatments of nuclear isomer transitions in weakly interacting massive particle (WIMP) models are given in Ref. ~\cite{Smirnov:2024jvj}.

\subsection{Internal conversion}
For the nuclear cascade we treat the $9^- \!\to\! 2^+$ IC and the subsequent $2^+\!\to\!1^+$ $\gamma$ emission or IC as a single prompt event. 
With Monte Carlo simulations based on Geant4~\cite{Agostinelli:2003,Allison:2006,Allison:2016}, we estimate the total deposited energy in the absorber and then fold the energy resolution. 
As shown in Fig.\ref{fig:prompt} (top), two peaks appear at $77$~keV (full de-excitation energy including IC electrons, $\gamma$ rays, and atomic X rays) and $37.7$~keV (energy deposit from $9^- \!\to\! 2^+$ step only; the subsequent $2^+ \!\to\! 1^+$ de-excitation escapes). 
Unless otherwise stated, we define the prompt region of interest~(ROI) as $E_{\rm peak}\pm 2\Delta E_\sigma$, with $\Delta E_\sigma=\Delta E_\mathrm{FWHM}/(2\sqrt{2\ln 2})$. The fraction of true IC events whose prompt energy falls inside the ROI is $\,\varepsilon^{\mathrm{IC}}_{p}=0.944$. 

\subsection{Strongly interacting DM de-excitation}
In an underground laboratory, overburden induces multiple scatterings for a strongly interacting DM component, effectively reducing the incident speed by a factor $\eta^{-1}$ while enhancing the number density by $\eta$~\cite{Pospelov:2019vuf}. 
For $\eta\gtrsim\mathcal{O}(10)$, the nuclear recoil from the DM-induced de-excitation is well approximated by
\begin{equation}
E_R \simeq \frac{\mu_{\chi N}}{m_N}\,\Delta_N,
\end{equation}
where $m_N$ is the target mass and $\mu_{\chi N}$ is the DM~($\chi$)–nucleus reduced mass. 
We define the nuclear energy gap as $\Delta_N \equiv E_i - E_f$, 
i.e., the energy released by the nuclear transition from the initial isomeric state to the final state. 
For $^{180\mathrm{m}}\mathrm{Ta}$, $\Delta_N = 37.7~\mathrm{keV} (9^-\!\to\!2^+$) and 
$\Delta_N = 77.2~\mathrm{keV} (9^-\!\to\!1^+$). 
For $m_\chi=1~\mathrm{TeV}$, we obtain $E_R(9^-\!\to\!2^+) \approx 32.3~\mathrm{keV}$ and $E_R(9^-\!\to\!1^+) \approx 66.1~\mathrm{keV}$. 

The deposited prompt energy therefore shows 
\begin{itemize}
  \item $9^-\!\to\!2^+$: a peak at $E_R+39.5~\mathrm{keV}$ arising from the $2^+\!\to\!1^+$  de-excitation, which includes $\gamma$ emission and IC electron, 
  \item $9^-\!\to\!1^+$: a peak at $E_R$ (no additional $\gamma$/electron in the prompt).
\end{itemize}
The event (counting) rate $R$ can be written as
\begin{equation}
 R= \frac{N_T (f\eta)\rho_\chi}{m_\chi}\frac{\mu_{\chi N}}{q_0}\sigma_n S_f(q), 
\end{equation}
where $q_0=\sqrt{2m_N \Delta_N}$. 
The nuclear-structure factor follows Ref.~\cite{Pospelov:2019vuf}:
\begin{align}
S_f(q) &\simeq \sum_{L} j_L^2(qR)\;\varepsilon_H(J_i,L,\Delta K),
\\
\varepsilon_H &= \big[M_0(E L)\big]^2\,\big[F_\gamma^{\,\Delta K-L}\big]^2 ,
\end{align}
with $R=r_0 A^{1/3}$, $r_0\simeq1.24$~fm, $M_0\simeq0.35$, and $F_\gamma\simeq0.16$ for Ta, following \cite{Ejiri:2017dro}. 

With $\pm2\sigma$ ROIs we obtain prompt efficiencies $\varepsilon^{\mathrm{sDM},2^+}_{p}=0.84$ and $\varepsilon^{\mathrm{sDM},1^+}_{p}=0.95$ (Fig.~\ref{fig:prompt}, middle). 
Here, the subscript “sDM” denotes the strongly interacting DM scenario. 

\subsection{Inelastic DM de-excitation}
\begin{figure}[t]
\includegraphics{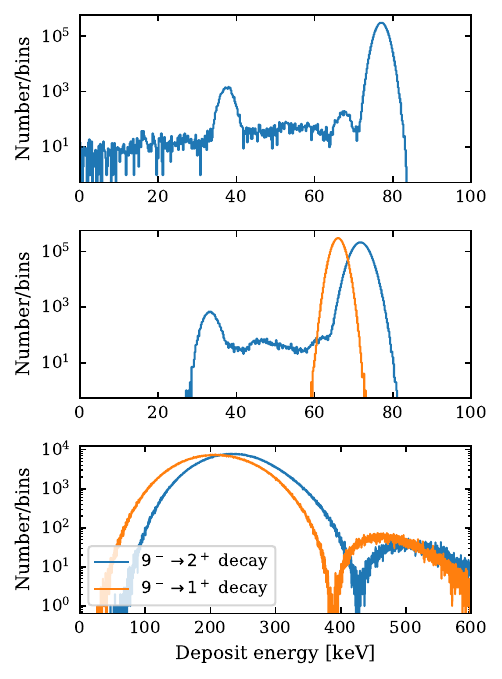}
\caption{\label{fig:prompt}
Deposited–energy distributions in a single Ta absorber for (top) internal conversion (IC), (middle) strongly interacting DM–induced de–excitation, and (bottom) inelastic DM–induced de–excitation. The spectra are obtained from Geant4 simulations including full containment of low–energy secondaries and folded with the assumed TES energy resolution. The middle and bottom panels are shown for a DM mass $m_\chi =1$~TeV; for inelastic DM we take a representative mass splitting $\Delta m = 50$~keV. }
\end{figure}

We introduce a mass splitting $\Delta m$ for inelastic DM and use the standard exo/endothermic kinematics for the minimum speed,
\begin{equation}
v_{\min}(E_R)=\frac{\big|\, m_N E_R/\mu_{\chi N} - \Delta E\,\big|}{\sqrt{2m_N E_R}}, 
\end{equation}
where $\Delta E \equiv \Delta_N - \Delta m$, together with a truncated Maxwellian halo to compute the mean inverse speed $\eta(v_{\min})=\int_{v > v_{\min}}d^3v f(v)/v$. 
Here $f(v)$ is the normalized DM velocity distribution in the Earth frame.
The differential recoil spectrum reads
\begin{align}
\frac{dR}{dE_R}
= N_T\frac{\rho_\chi}{m_\chi}\;
\frac{2m_N\sigma_n}{\mu_{\chi n}^2}\;
S_f(q)\,\eta(v_{\min}) \, .
\end{align}
where $q=\sqrt{2m_NE_R}$ and the DM~($\chi$)–nucleon reduced mass $\mu_{\chi n}$. 
For $9^-\!\to\!2^+$ we treat the $2^+$ level as a member of the $K^\pi=1^+$ ground band, giving $\Delta K=|9-1|=8$; the dominant multipole is $L=7$ with one unit of $K$-forbiddenness ($\nu=\Delta K-L=1$), while $L=8$ is subdominant. For $9^-\!\to\!1^+$ we take $\Delta K=8$ and $L=8$ (no $K$ penalty). The deposited prompt spectrum is shown in Fig.~\ref{fig:prompt} (bottom). Taking into account the characteristic shape of the inelastic recoil spectrum, we adopt a fixed $\pm 20$~keV window around the peak as a compromise between signal efficiency and background rejection; this choice is close to optimal for the resolutions considered and is robust against small variations in the window size. For $m_\chi=1$~TeV and $\Delta m=50$~keV, we find $\varepsilon^{\mathrm{iDM},2^+}_{p}\simeq0.27$ and $\varepsilon^{\mathrm{iDM},1^+}_{p}\simeq0.26$. Here, the subscript “iDM” denotes the inelastic DM. 

\subsection{Electron capture}
\begin{figure}[t]
\includegraphics{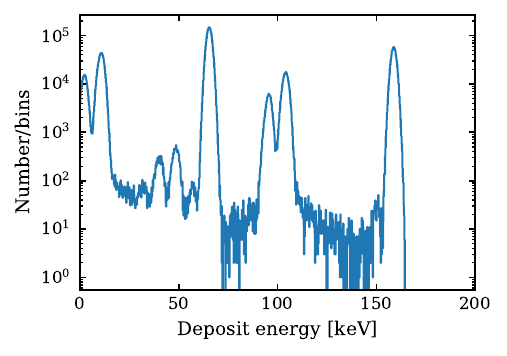}
\caption{\label{fig:delayed}Deposited–energy distribution for the delayed EC channel in a single Ta absorber. The spectrum shows a prominent peak at the K–shell binding energy $E_K\simeq65.35~\mathrm{keV}$ and, when the de–excitation $\gamma$ ray of energy ($E_{\gamma}=93.3~\mathrm{keV}$) is fully contained, a second peak at $E_K + E_{\gamma} \simeq 158.7~\mathrm{keV}$}
\end{figure}

We take EC of $^{180}$Ta to $^{180}$Hf as the delayed events, following the ENSDF evaluation for $A=180$~\cite{McCutchan2015A180}. 
The total EC branching fraction is $86(3)\%$, with the remaining $14(3)\%$ proceeding via $\beta^{-}$ decay; the latter constitutes an absolute inefficiency for the delayed tag. 
Conditional on EC, we assume that $71\%$ of decays populate the $0^+$ ground state of $^{180}$Hf and $29\%$ populate the $2^+$ state, as given in the same evaluation~\cite{McCutchan2015A180}.
Including the subsequent $2^+\!\to 0^+$  transition at \(E_{\gamma}=93.3~\mathrm{keV}\), the delayed spectrum (Fig.~\ref{fig:delayed}) exhibits a prominent peak at the $K$-shell binding energy \(E_K\simeq65.35~\mathrm{keV}\) and, when the $\gamma$ ray is fully contained, a second peak at $E_K + E_{\gamma} \simeq 158.7~\mathrm{keV}$. 
For the delayed selection, we define ROIs as $[\,E_K \pm 2\Delta E_\sigma\,]$ and $[\,E_K+E_{\gamma} \pm 2\Delta E_\sigma\,]$. The resulting delayed efficiency is $\varepsilon_{d}\simeq0.65$.

\subsection{Delayed coincidence selection}
The expected number of detected events in the delayed-coincidence analysis is
\begin{equation}
N_S^{Z} \;=\; \frac{\ln 2}{T_{1/2}}\; N_{\mathrm{TES}}\; N_T\; T_{\mathrm{obs}}\;
\varepsilon^{Z}_{p}\; \varepsilon_d\; \varepsilon_{\mathrm{EC}}\; \varepsilon_{dT}\,,
\end{equation}
where $Z\in\{\mathrm{IC}, (\mathrm{sDM},1^+), (\mathrm{sDM},2^+),(\mathrm{iDM},1^+), (\mathrm{iDM},2^+)\}$; $T_{1/2}$ is the relevant half-life; $N_{\mathrm{TES}}$ is the number of pixels; $N_T$ is the number of $^{180\mathrm{m}}$Ta nuclei per pixel; $T_{\mathrm{obs}}$ is the observation time; and $\varepsilon_{dT}$ is the delayed-window acceptance. 
In this work we adopt a delayed-coincidence window corresponding to three
half-lives of the EC decay of $^{180}$Ta, $\Delta t \equiv 3\,T_{1/2}(^{180}\mathrm{Ta}) \simeq 24.5~\mathrm{h}$,
which yields $\varepsilon_{dT}=0.875$. For a cubic Ta absorber of edge 1.5~mm and the natural isomeric fraction $1.2\times10^{-4}$, we estimate $N_{T}\simeq2.3\times10^{16}$ nuclei per pixel.

\section{Background model}
\begin{figure}[t]
\includegraphics{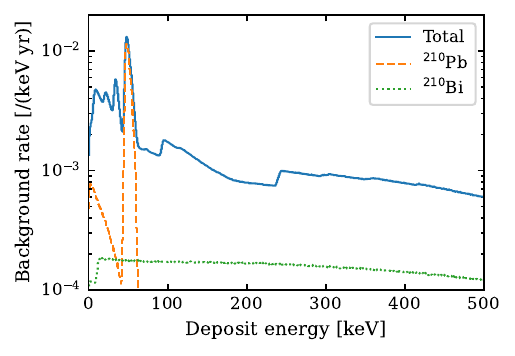}
\caption{\label{fig:background}Simulated background energy spectra in the Ta absorber for the intrinsic $^{238}$U and $^{232}$Th contaminations assumed in this work. The contributions from $^{210}$Pb and $^{210}$Bi are also shown; the sequential $^{210}$Pb\,$\to$\,$^{210}$Bi decay constitutes a correlated background that can pass the delayed-coincidence selection. 
}
\end{figure}

Assuming that external radiation is fully mitigated by appropriate shielding and that measurements are performed underground where the residual cosmic-ray flux is negligible, the dominant background is intrinsic radioactivity in the Ta absorber. Adopting radiopurity levels demonstrated in prior studies—$^{238}$U: 5~ppt and $^{232}$Th 10~ppt~\cite{Cerroni:2023qoo}, 
and assuming secular equilibrium throughout the decay chains, the deposited-energy spectrum from intrinsic backgrounds is shown schematically in Fig.~\ref{fig:background}.  
The structure around 50~keV is attributable to the $\beta$ decay of $^{210}$Pb, 
while features below 40~keV arise primarily from the $\beta$ decay of $^{228}$Ra. The estimated single-pixel background rates in each ROI are: IC, $n^{\mathrm{IC}}_{p}=7.6\times10^{-3}\ \mathrm{yr}^{-1}$; strongly interacting DM, $n^{\mathrm{sDM},2^+}_{p}=8.0\times10^{-3}\ \mathrm{yr}^{-1}$ and $n^{\mathrm{sDM},1^+}_{p}=8.1\times10^{-3}\ \mathrm{yr}^{-1}$; delayed channel, $n_{d}=1.4\times10^{-2}\ \mathrm{yr}^{-1}$. Here $n_p^Z$ ($Z\in\{\mathrm{IC},\,\mathrm{sDM},\,\mathrm{iDM}\}$) denotes the
total single-pixel background rate integrated over the prompt ROI for channel $Z$,
and $n_d$ is the corresponding rate integrated over the delayed ROI.

The expected number of background events $N_B^Z$ in channel $Z$ after the
delayed-coincidence selection is given by the sum of two components: the number of
accidental coincidences between uncorrelated prompt and delayed events,
$N_{\mathrm{ACC}}^Z$, and the number of correlated background events from the
sequential $^{210}\mathrm{Pb}\to{}^{210}\mathrm{Bi}$ decay within the Ta absorber,
$N_{210\mathrm{Pb}}^Z$:
\begin{equation}
  N_B^Z \;=\; N_{\mathrm{ACC}}^Z + N_{210\mathrm{Pb}}^Z.
  \label{eq:NB}
\end{equation}

Assuming independence across pixels, the accidental contribution $N_{\mathrm{ACC}}^Z$ can be written as 
\begin{equation}
N^{Z}_{\mathrm{ACC}}\;=\; N_{\mathrm{TES}}\, T_{\mathrm{obs}}\, n_{p}^{Z}\, n_{d}\,
\frac{\Delta t}{8760}, 
\end{equation}
where $\Delta t =24.5~\mathrm{h}$ is the delayed-coincidence window in hours. 
The factor $1/8760$ converts the window width to a fraction of a year
when $T_{\mathrm{obs}}$ is expressed in years. 

The sequential decay $^{210}\mathrm{Pb}\to{}^{210}\mathrm{Bi}$ can also produce fake
delayed-coincidence events.
$^{210}$Pb undergoes $\beta^-$ decay with endpoint energy
$Q_\beta = 63.5~\mathrm{keV}$ to $^{210}$Bi (see Fig.~\ref{fig:background}).
The daughter $^{210}$Bi then $\beta$-decays with a half-life of
$T_{1/2} \simeq 5.0~\mathrm{d}$ and $Q\simeq 1.16~\mathrm{MeV}$ to $^{210}$Po.
A $^{210}$Pb decay can provide the prompt energy deposit in the ROI for channel $Z$,
while the subsequent $^{210}$Bi decay may fall in the delayed EC energy window.
From our spectral simulation and detector response, the probability that the
$^{210}$Bi $\beta$ decay deposits energy in the delayed ROI is
$\epsilon_{\mathrm{Bi}} = 0.017$, and the probability that this $^{210}$Bi decay
occurs within the delayed-coincidence time window is $\epsilon_t = 0.131$.
\begin{equation}
N^{Z}_{^{210}\mathrm{Pb}} \; =\; 
N_{\mathrm{TES}}\,  T_{\mathrm{obs}}\, n_{210\mathrm{Pb}}^{Z}\, \epsilon_{\mathrm{Bi}}\, \epsilon_{t}, 
\end{equation}
where $n_{210\mathrm{Pb}}^Z$ is the single-pixel rate in the prompt ROI for channel $Z$ arising specifically from $^{210}$Pb $\beta$ decays.

\section{Discovery Power}
Following Ref.~\cite{Agostini:2017jim}, we estimate the number of signal events $N^Z_S$ corresponding to a 3$\sigma$ discovery for an expected background $N^Z_B$ with the Asimov approximation~\cite{Cowan:2010js}.  
We then convert $N^Z_S$ to a constraint on the half-life ($T_{1/2}$) or, for DM channels, on the interaction cross section. 

\begin{figure}[t]
\includegraphics{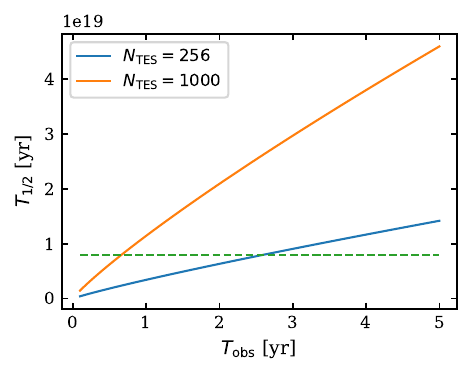}
\caption{\label{fig:ic}Half-life $T_{1/2}$ at the $3\sigma$ discovery level as a function of observation time $T_{\mathrm{obs}}$ with $N_{\mathrm{TES}}=256$ and $N_{\mathrm{TES}}=1,000$. The dashed horizontal line indicates the theoretical expectation for the IC half–life of $^{180\mathrm{m}}$Ta~\cite{Ejiri:2017dro}. }
\end{figure}

As a first benchmark, we consider the IC channel with arrays of $N_{\mathrm{TES}}=256$ and 1,000 pixels. Figure \ref{fig:ic} shows the $3\sigma$ discovery reach in half-life as a function of observation time; under these assumptions, the sensitivity reaches the theoretical value~($8\times 10^{18}~\mathrm{yr}$~\cite{Ejiri:2017dro}) after 2.6 yr for $N_{\mathrm{TES}}=256$ pixels and 0.66 yr for $N_{\mathrm{TES}}=1,000$. 

\begin{figure}[t]
\includegraphics{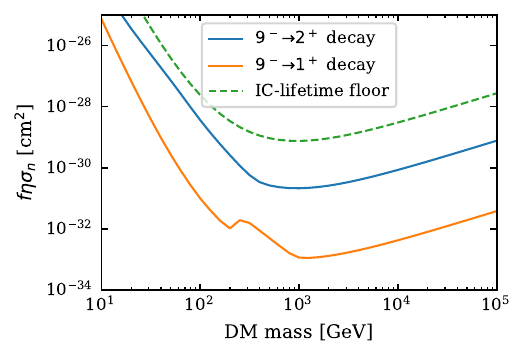}
\caption{\label{fig:sDM}$3\sigma$ discovery reach in the combination  $f\,\eta\,\sigma_n$ as a function of DM mass $m_\chi$ for strongly interacting DM. We assume an array with $N_{\mathrm{TES}}=10,000$ and five years of observation. Dashed curves indicate the $\gamma$-only IC-lifetime floor (see text).
}
\end{figure}

\begin{figure}[t]
\includegraphics{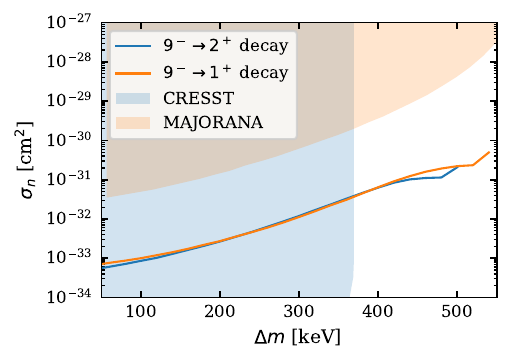}
\caption{\label{fig:iDM}$3\sigma$ discovery reach in the DM–nucleon cross section $\sigma_n$ as a function of the mass splitting $\Delta m$ for $m_\chi=1$~TeV DM. We assume an array with $N_{\mathrm{TES}}=10,000$ and five years of observation. 
The shaded ``MAJORANA'' region shows the official inelastic-DM constraint reported by the MAJORANA Collaboration~\cite{Majorana:2023ecz}, while the ``CRESST'' region is included for qualitative comparison only and follows the model-dependent reinterpretation of Ref.~\cite{Bramante:2016rdh}. }
\end{figure}

For the DM search, Figs.~\ref{fig:sDM} and \ref{fig:iDM} summarize the five-year discovery reach in cross section for an array with $N_{\mathrm{TES}}=10,000$. 
For strongly interacting DM, the reach depends on a site-dependent factor $\eta$ and on the fraction $f$ of the total DM density residing in the strongly interacting component. Accordingly, we present the discovery reach in terms of $f\,\eta\,\sigma$. 
In our benchmark configuration, the correlated background from the sequential $^{210}\mathrm{Pb}\rightarrow{}^{210}\mathrm{Bi}$ decay, $N^{Z}_{210\mathrm{Pb}}$, dominates the total background in the mass range $m_\chi \simeq 300$--$600~\mathrm{GeV}$ for the $9^- \rightarrow 1^+$ transition and in $m_\chi \simeq 20$--$600~\mathrm{GeV}$ for the $9^- \rightarrow 2^+$ transition. 
In $\gamma$-only HPGe measurements, IC-induced and DM-induced de-excitations are observationally indistinguishable on an event-by-event basis. As a result, for the $9^- \rightarrow 2^+$ branch the inferred DM sensitivity saturates once the DM-induced rate falls below the IC de-excitation rate; this $\gamma$-only ``IC-lifetime floor'' is indicated by the dashed curves in Fig.~\ref{fig:sDM}. 
Moreover, for the $9^- \rightarrow 1^+$ branch a $\gamma$-only HPGe search provides little or no prompt handle, since the DM-induced transition need not produce an observable prompt $\gamma$ ray (and the accompanying nuclear recoil is not measurable in a standard external-source HPGe configuration). In our $\gamma$-TES approach, by contrast, the absorber is both the source and detector and can measure the nuclear-recoil energy associated with the de-excitation; for the $9^- \rightarrow 1^+$ channel this can be complemented by a delayed-coincidence tag from the subsequent $^{180}$Ta ground-state decay (EC/$\beta$, $T_{1/2}\simeq 8.15$~h).
While the reach depends on the site, the $\gamma$-TES-based search indicates a distinctive discovery opportunity for strongly interacting DM.

Figure~\ref{fig:iDM} indicates that, for inelastic DM---analogous to the strongly interacting case---our approach can extend beyond the bounds set by $\gamma$-only measurements of $^{180\mathrm{m}}$Ta decay and probe regions of parameter space that are not yet accessible to conventional direct-detection experiments.
In particular, the $^{180\mathrm{m}}$Ta de-excitation channel retains sensitivity to mass
splittings of order several hundred keV, extending the reach to larger $\Delta m$ than the
CRESST region shown in Fig.~\ref{fig:iDM}.
This complementarity arises because our search targets the de-excitation energy deposited in the absorber (and, where applicable, a delayed EC tag), rather than relying solely on nuclear-recoil signatures as in conventional direct-detection analyses.
For clarity, the CRESST region is taken from the reinterpretation in Ref.~\cite{Bramante:2016rdh} and is included for qualitative comparison only.
It is model- and analysis-dependent and should not be viewed as an official CRESST inelastic-DM limit.

\section{Summary}

We have presented a quantitative study of the discovery potential of $\gamma$-TES arrays with Ta absorbers for the de-excitation of $^{180\mathrm{m}}$Ta. By operating the detector in a source-equals-detector configuration, we exploit two capabilities absent in conventional HPGe searches: (i) near-unity containment of low-energy secondaries (IC electrons and characteristic X rays) and of the nuclear recoil, enabling calorimetric, event-by-event measurements of the total deposited energy and thereby allowing IC- and DM-induced channels to be discriminated based on their energy distributions; and (ii) a delayed-coincidence tag using the subsequent electron capture (EC) decay of $^{180}$Ta to $^{180}$Hf, which is directly visible via the atomic relaxation energy in the daughter (characteristic X rays and Auger electrons), independent of whether the EC populates the ground or an excited state.

For the IC channel, we find that arrays with $N_{\mathrm{TES}} \simeq 10^2$--$10^3$ pixels and realistic radiopurity can reach the theoretically expected IC half-life of $^{180\mathrm{m}}$Ta within observation times of order a few years. Thus, a medium-scale $\gamma$-TES array provides a realistic path to either observe the long-sought decay of $^{180\mathrm{m}}$Ta or to significantly strengthen existing half-life limits. 

For DM-induced de-excitation, we have examined two benchmark scenarios: a strongly interacting DM component parametrized by $f\,\eta\,\sigma_n$, and inelastic DM characterized by a mass splitting $\Delta m$. Assuming an array with $N_{\mathrm{TES}} = 10^4$ pixels and a five-year observation, the projected $3\sigma$ discovery reach for strongly interacting DM depends on the site-dependent factor $\eta$ but can exceed bounds implied by HPGe measurements, which are ultimately limited by the unknown IC lifetime. For inelastic DM, the ability to detect the nuclear recoil spectrum associated with de-excitation allows the $\gamma$-TES-based search to probe regions of $(\Delta m,\sigma_n)$ parameter space that are not yet accessible to current direct-detection experiments, while remaining consistent with existing limits from CRESST and MAJORANA.

In our DM sensitivity estimates, we do not include the theoretically predicted IC rate as a fixed background, given that it has not yet been measured experimentally.
Instead, we emphasize that a direct measurement of the $9^-\!\to 2^+$ IC half-life in $^{180\mathrm{m}}$Ta is an important complementary goal: such a measurement would provide a strong empirical calibration of the nuclear-structure inputs entering our structure factor (including the overall transition strength and the K-hindrance encoded in parameters such as $M_0$ and $F_\gamma$), and would thereby substantially reduce the associated theoretical uncertainties.

The dominant systematic uncertainties in our projections arise from these nuclear-structure inputs, in particular the reduced matrix element $M_0$ and the hindrance factor $F_\gamma$, as well as from assumptions about intrinsic contamination in the Ta absorber. A precise IC half-life measurement would therefore largely control the nuclear-structure component of this uncertainty.

On the experimental side, a near-term program includes a prototype multi-pixel run to demonstrate delayed-coincidence selection, validate background modeling, and refine radiopurity requirements. As part of this effort, we performed an HPGe assay of a specially procured Ta sample using a dedicated low-background detector installed in the Kamioka underground laboratory~\cite{Ichimura:2023pxx}. No statistically significant activity attributable to the U and Th decay chains was observed. 
Further progress in constraining the background will ultimately require direct measurements with the actual materials and experimental configuration.

With these developments, $\gamma$-TES arrays with Ta absorbers offer a compelling and complementary avenue to explore isomer de-excitation physics and to search for DM-induced transitions in $^{180\mathrm{m}}$Ta. 
Several of the authors are currently installing and commissioning a dilution refrigerator in the Kamioka underground laboratory, with the long-term goal of deploying a $\gamma$-TES array there to carry out a high-sensitivity search for the decay of $^{180\mathrm{m}}$Ta.

\begin{acknowledgments}
This work was supported by JSPS KAKENHI Grant Numbers 22K18709, 22H04946, 23K03415, 24H00209, 24H00225, 24H02237, and 24H02244; Matching fund between Tohoku University and National Institute of Advanced Industrial Science and Technology (AIST) in 2023, 2024; and the Grant for Basic Science Research Projects from The Sumitomo Foundation (Grant No. 2502590). We acknowledge JX Advanced Metals Corporation for providing the special sample. 
\end{acknowledgments}

\section*{Data availability}
The data are not publicly available. The data are available from the authors upon reasonable request.
 
\input{ref.bbl}
\end{document}

%% file: ref.bbl
\providecommand{\noopsort}[1]{}\providecommand{\singleletter}[1]{#1}%